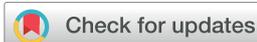

# Modelling structure and ionic diffusion in a class of ionic liquid crystal-based solid electrolytes†

Md Sharif Khan, [ID] *[a] Ambroise Van Roekeghem,[a] Stefano Mossa, [ID] [b] Flavien Ivol,[a] Laurent Bernard,[a] Lionel Picard[a] and Natalio Mingo [ID] *[a]

Next-generation high-efficiency Li-ion batteries require an electrolyte that is both safe and thermally stable. A possible choice for high performance all-solid-state Li-ion batteries is a liquid crystal, which possesses properties in-between crystalline solids and isotropic liquids. By employing molecular dynamics simulations together with various experimental techniques, we have designed and analyzed a novel liquid crystal electrolyte composed of rigid naphthalene-based moieties as mesogenic units, grafted to flexible alkyl chains of different lengths. We have synthesized novel highly ordered lamellar phase liquid crystal electrolytes at 99% purity and have evaluated the effect of alkyl chain length variation on ionic conduction. We find that the conductivity of the liquid crystal electrolytes is directly dependent on the extent of the nanochannels formed by molecule self-organization, which itself depends non-monotonously on the size of the alkyl chains. In addition, we show that the ion pair interaction between the anionic center of the liquid crystal molecules and the $Li^+$ ions plays a crucial role in the overall conductivity. Based on our results, we suggest that further improvement of the ionic conductivity performance is possible, making this novel family of liquid crystal electrolytes a promising option for the design of entirely solid-state $Li^+$ ion batteries.

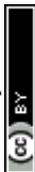

## 1. Introduction

Global warming, pollution and natural disasters are present main challenges for mankind, and all are directly related to the uncontrolled use of fossil fuels. Development of alternative transport systems such as electric vehicles is consequently crucial, calling for significant improvements in battery technologies, including enhanced electrochemical efficiency, cyclability, environmental friendliness and low costs.[1,2] In particular, in order to increase the energy density, it is required to develop new electrolytes and electrode materials.[3,4] Compared to all other existing secondary batteries, Li-ion batteries (LIBs) exhibit the highest volumetric and gravimetric energy density.[5] The standard electrolytes employed in commercial LIBs are typically a solution of lithium hexafluorophosphate dissolved in a binary or tertiary mixture of organic carbonates such as ethylene carbonate, dimethyl carbonate and ethyl methyl carbonate, and additives.[6,7] Unfortunately, such organic electrolytes are highly toxic and flammable, and are prone to dendrite growth, with a limited electrochemical potential window.[8,9]

Solid state electrolytes have a potential to overcome these limitations, along with a few new battery chemistries.[5,10] For instance, crystalline ceramics with sufficient cation conductivity have been described.[11–14] However, these solid electrolytes have poor contacts with the electrodes due to their intrinsic stiffness, are brittle in nature, and have low cyclability.[15,16] In contrast, owing to their good mechanical properties and flexibility, satisfying interaction with electrodes, and ability to compensate electrode volume changes, solid state polymer electrolytes stand out as a promising choice.[17–19] Poly-ethylene oxide (PEO) is one of the most widely studied polymer electrolytes, with high salt solubility without phase separation, and a particularly low glass transition temperature.[20–22] Yet, despite the good ion solubility, PEO has low cation transference number and mobility, due to the strong coordination between the cation and multiple oxygen atoms comprised in its structure.[23–25] The lower the cation transport number, the greater the role played by the anion in determining the ionic conductivity, which influences the charge and discharge processes, and the formation of dendrites[26,27] drastically reducing battery performance.

[a] Université Grenoble Alpes, CEA, LITEN, 17 rue des Martyrs, 38054 Grenoble Cedex 9, France. E-mail: sharifkhanjnu@gmail.com, natalio.mingo@cea.fr
[b] Université Grenoble Alpes, CEA, IRIG-MEM, 17 rue des Martyrs, 38054 Grenoble Cedex 9, France

† Electronic supplementary information (ESI) available: Molecular synthesis scheme and NMR characterization; Impedance Spectroscopy; TGA curves of the LC molecules; ball and stick structure of the typical LC molecule; per-atom partial charge distribution; volume and density changes of LC molecules as a function of temperature; RDF at different temperatures; summary of the RDF peak; static structural factor; Onsager transport coefficients and the transport mechanism of $Li^+$ in BS-Li-8 and BS-Li-16. See DOI: https://doi.org/10.1039/d3cp05048c

Increasing the cation transport number by anchoring the anion is a possible option to improve the performance of the electrolyte. In this direction, liquid crystal (LC) electrolytes have recently attracted significant attention, due to the presence of organized ion conduction pathways, high transference number, wider potential and temperature windows, low flammability, and cost-effective bulk production.[28–31] Ionic conductivities of LC electrolytes in the range of $10^{-6}$ to $10^{-3}$ S cm$^{-1}$ at room temperature have been reported[32],[33]. Sakuda *et al.* demonstrated reversible charge–discharge cycling with different cathodes and Li-metal anodes.[33] Ahmed *et al.* showed significant dendrite growth suppression on Li-metal anodes by LC electrolytes, by using phase field models and density functional theory calculations.[34] Despite the possible important impact of these electrolytes on the LIB performance, very few materials of this class have been thoroughly analyzed. As a consequence, further development of these materials is limited due to the lack of a sufficient fundamental understanding of the ion transport mechanism.

The goal of this article is to address this issue. We have designed and synthesized novel LC organic electrolytes for LIBs, with a naphthalene mesogenic moiety bearing a lithium sulfonate group and connected to two flexible long-alkyl chains of variable length. We have employed molecular dynamics simulations together with various experimental techniques for a comprehensive understanding of both the bulk structure and the transport mechanism. We shed new light on the influence of both the non-polar alkyl chain structure and temperature on the bulk structural arrangement and conductivity of such electrolytes.

## 2. Molecular concept

The possible design space of LC electrolytes is vast. A complete understanding of the structure–property relationships, molecular mobility, and ionic conductivity mechanisms is therefore crucial to tailor and optimize their properties for energy storage applications. When the anionic moiety is anchored to the LC, the self-assembly of these compounds is driven by the strongest ionic forces, followed by the van der Waals interactions. The degree of freedom that one can tune to control the mesomorphic nature is the alkyl chain morphology, while the directionality of ordering is mainly controlled by the mesogenic part.

Here, inspired by our previous work,[35] we have considered LCs based on a lithium salt moiety directly anchored on an unusual mesogenic group attached to long flexible chains of variable lengths. More concretely, they consist of an aromatic naphthalene-based core with a sulfonate anion and alkyl chains attached to the core *via* a tertiary amine (see Fig. 1). This particular choice has been dictated by the hypothesis that the combination of the flexible long-chains and the mesogen-like naphthalene sulfonate unit would induce the self-assembly of nanostructures in well-defined one-, two- or three-dimensional pathways.[35] This would in turn enhance the lithium conductivity compared to a typical polymer electrolyte, where the ions are randomly dissolved and dissociated in the polymer chains. In this case, the anionic centre is covalently attached to the naphthalene group, and ideally only the cation contributes to the ionic conductivity. For the selected LC molecules, properties can be tuned through the mesogenic moiety structure, length of the spacer and functional end groups, and cationic charge carrier.

We emphasize that in order to maximize the ionic conduction in the direction perpendicular to the surface of the electrodes, a parallel orientation of the ion-conducting paths of different macroscopic domains is desired. Typically, the dipole–dipole interactions between the alkyl chains of distinct molecules are the main factor determining the molecular structure. Also, it is well known that mesomorphism appears for alkyl chains containing a minimum of six carbon atoms.[36] If the carbon chain length is below this threshold, with increasing temperature the system evolves from a solid crystalline state directly to an isotropic liquid phase. It should also be noted that with an increase in the alkyl chain length, the distance between ionic domains also increases.

## 3. Experiments and simulation methods

**Experiments**

4-Amino-1-naphthalenesulfonic acid (Sigma Aldrich, 97%), *N,N*-diisopropylethylamine (Sigma Aldrich, >99%, stored in an Ar-filled glovebox), DMF (Sigma Aldrich, anhydrous 99.8%, stored in an Ar-filled glove box), 1-iodododecane (Sigma Aldrich, 98%), 1-iodohexadecane (Sigma Aldrich, 95%), 1-iodooctane (Sigma Aldrich, 98%), methanol (Sigma Aldrich, 99.8%), and lithium hydroxide (Sigma Aldrich, 99.9%) were used as received and used without further purification. All reactions were carried out under an argon atmosphere in anhydrous solvents.

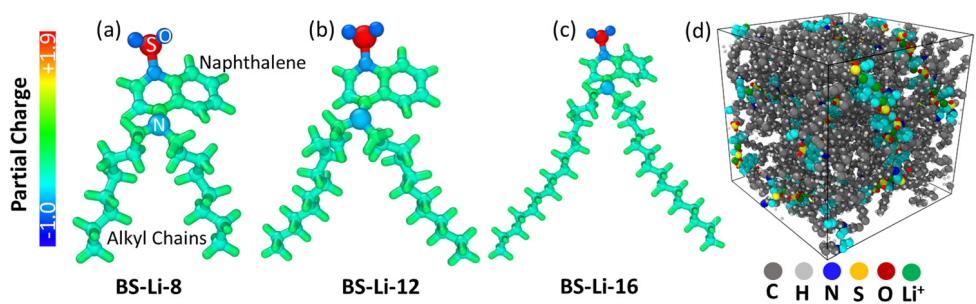

Fig. 1 Ideal structures and distributions of the atomic partial charge of the LC molecules: (a) BS-Li-8, (b) BS-Li-12, and (c) BS-Li-16. (d) Typical simulation system (The turquoise color denotes the carbon atoms belonging to naphthalene rings).

$^1$H, $^7$Li and $^{13}$C nuclear magnetic resonance (NMR) spectra were collected at 25 °C using a Bruker ADVANCE 400 MHz spectrometer, for samples solubilized in deuterated solvents (Fig. S2–S4, ESI†). Chemical shifts of $^1$H, $^7$Li and $^{13}$C NMR signals are reported using dimethyl sulfoxide – d6 ($\delta$ = 2.50 ppm) as the internal standard, and expressed by chemical shifts in parts per million ($\delta$ = ±0.01 ppm), multiplicity (s (singlet), d (doublet), t (triplet), and m (multiplet)), coupling constant (Hz = ±0.3 Hz), and relative intensity. Details pertaining to the synthesis and NMR measurements of the LCs are provided in the ESI.†

Polarized optical microscopy (POM) studies were performed using an Eclipse LV100ND microscope, working in both transmission and reflection modes, equipped with a series of ×2.5, ×5, ×10, ×20 and ×50 objectives and a Linkam heating stage.

Small-angle X-ray scattering (SAXS) data were measured in transmission mode using a home-made SAXS camera: a Bruker–Nonius (FR591) rotating anode generator with Cu Kα radiation ($\lambda$ = 1.5418 Å) at 45 kV and 66 mA, a gas detector 2D (Xe/CO$_2$) with a 20 cm/20 cm window (resolution 500 μm) and vacuum scattering tubes adjustable to several lengths. In this study, a distance of 42 cm was used between the sample and the detector. The distance calibration was performed by using silver behenate as the reference sample. The samples were dried overnight at 80 °C prior to characterization and were placed in quartz sealed capillary during the characterization. Temperature and heating of the capillary was controlled using a home-made device (±1 °C). The SAXS profiles were obtained by reduction of the two-dimensional data by radial integration of the intensity, after data correction for the background scattering from the empty cell.

Thermal stability of the materials was assessed using a TGA TA-550 instrument and the measurements were performed under a He-atmosphere at 10 °C min$^{-1}$.

To evaluate the ionic conductivity, $\sigma$, electrochemical impedance spectroscopy (EIS) was carried out at various temperatures, collecting data in the range 303 K to 353 K, at intervals of 10 K. Based on these data and knowing the cell constants (distance between the electrodes, $t$, and cross-sectional area of the sample, $A$), we next evaluated the ionic conductivity for each sample using the following procedure. We obtained the resistance of the cables, $R1$, the resistance of the electrolyte, $R2$, (typically ranging from 5000 to 20 000 ohms), and a constant phase element, $Q2$, by adjusting the Nyquist plots to the equivalent circuit formula, $Z = R1 + Q2/R2$, where $Z$ is the complex impedance (Fig. S5, ESI†). The conductivity was finally calculated at all temperatures from $\sigma = t \cdot (A \cdot R2)^{-1}$. The ionic conductivity data plotted against the inverse temperature $(10^3/T)$ were next fitted by linear regression to the logarithmic form of the Arrhenius equation, in order to better qualify the nature of the ionic process(es) involved.

### Simulation

All-atoms molecular dynamics (MD) simulations were conducted using a large-scale atomic/molecular massively parallel simulator (LAMMPS) molecular dynamics code.[37] The OPLS-2005 force field was employed to represent the electrolytes and their partial charges.[38] To account for the well-recognized overestimation of the ion–ion interactions in non-polarizable force fields, the partial charges of all the ionic species were scaled by a factor of 0.9.[39] The particle–particle particle–mesh (PPPM) method[40] was used to compute long-range Coulomb interactions, with a real-space cut-off of 1.2 nm. We fixed the cutoff length for the van der Waals interaction to 1 nm, and the Lorentz–Berthelot rules were adopted for the mixed Lennard-Jones interactions. The NPT and NVT simulations were conducted with time steps of 0.5 and 1.0 fs, respectively. The Nosé–Hoover algorithm was used to control the temperature, while the SHAKE algorithm was employed to constrain the C–H bonds in the alkyl chains.[41,42]

We prepared different numerical samples as follows. 90 LC molecules were randomly inserted in a cubic simulation box periodic in all directions (Fig. 1), and the energies were next minimized by using a conjugate gradient method. The electrolytes were subsequently annealed from 100 to 900 K at 1 atm, in the NPT ensemble for 30 ns. Once the systems reached equilibrium at 900 K, the temperature was reduced gradually in steps of 10 K. At each step, 20 ns NPT-simulations were conducted to attain the system density corresponding to atmospheric pressure. Throughout the temperature reduction, electrolyte topologies were collected in the range 353 to 413 K, and used as the initial configuration for 30 ns of NVT simulations at all temperatures. System configurations were finally collected throughout this last production run and used for analysis.

## 4. Results

### 4.1 Structures

The structures of the considered LCs are shown in Fig. 1 and Fig. S1, S7 (ESI†). The macro-molecule contains a –SO$_3^-$ group as the negatively charged head, coupled to the alkyl chains (playing the role of flexible spacers) via a naphthalene group, which is likewise linked to the alkyl chains by a nitrogen atom. Depending on the considered alkyl chain lengths, i.e., the number (8, 12, 16) of alkyl groups in each chain, we have dubbed the molecules BS-Li-8 (lithium 4 – (dioctylamino) naphthalene – 1 sulfonate), BS-Li-12 (lithium 4 – (didodecylamino) naphthalene – 1 sulfonate), and BS-Li-16 (lithium 4 – (dihexadecylamino) naphthalene – 1 sulfonate), respectively. The fully relaxed isolated molecular structures for all cases together with the partial charge dispersions are depicted in Fig. 1(a)–(c). We used the 1.14*CM1A method to determine the partial charge distributions that we employed in the OPLSAA force field.[43] Prior to this method, we had also attempted to use a simpler semi-empirical QM model,[44,45] and also an *ab initio* DFT calculation using Gaussian (B3LYP with basis set STO-3G). Calculations using these two other methods did not reproduce the observed conductivity ordering, and were unable to provide a molecular level understanding of the macroscopic observations.

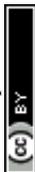

A few comments are in order. First, no important differences in the charge distributions can be observed in the three cases, as expected. Second, the partial charges of O and S in the $SO_3^-$ group are the highest negative and positive ones, respectively. As a result, the $SO_3^-$ group can be safely considered as the relevant anionic center of the LC molecules. The numerical values of the partial charge per atom are found in Table S1 (ESI†), and in Fig. 1(d) we show a typical simulation system snapshot containing 90 LC molecules. The different moieties are indicated with different colors, as detailed in the caption.

The working temperature range of the present LC electrolyte can be inferred from the variation of density (or, equivalently, volume) with temperature, as shown in Fig. S8 (ESI†). A drastic volume change is observed for all systems around 420–430 K, while no noticeable hysteresis is present when we decrease the temperature. On this basis, we conclude that the working temperature range for our electrolytes extends up to around 420 K.

Both the chemical composition and length of the alkyl chains have an important impact on the bulk structure and ion transport properties. Simulated partial radial distribution functions (RDFs) corresponding to specific atomic pairs for BS-Li-8, BS-Li-12, and BS-Li-16 are shown in Fig. 2(a) at $T$ = 413 K. RDFs for these same pairs at temperatures of 353, 373, and 393 K are shown in Fig. S9 (ESI†). The data in Fig. 2(a) show that, for all molecules, the $Li^+$ ion strongly binds to both S and O. This is not unexpected, as we already noticed that the partial charge distribution in Fig. 1 indicates $SO_3^-$ as the anionic center of these LC molecules. We note that $Li^+$ has a somewhat larger association degree with O and N in BS-Li-8, as indicated by the observation that the Li–O and Li–N first neighbor peaks are located at slightly closer distances than in the case of BS-Li-12 and BS-Li-16, including an additional tiny peak for the Li–N case at a very short distance of 0.23 nm, which is not observed for the other two molecules. Also, in the case of BS-Li-8, the Li–Li correlation peak appears at a rather longer distance than those for BS-Li-12 and BS-Li-16, whereas the S–O and S–S correlations are about the same for all cases. We conclude that while in BS-Li-8, $Li^+$ still strongly interacts with the negative sites of the anions, as expected, it remains a little farther apart from the LC molecules than in the case of BS-Li-12 and BS-Li-16. This difference makes the behavior of BS-Li-8 quite different from that observed in the other two cases which, in turn, look quite similar.

RDFs at $T$ = 353, 373, and 393 K show similar trends (see Fig. S9, ESI†). The neighboring distance values of the different correlations are summarized in Table S2 (ESI†). Careful visual inspection of the system configurations also conveys important information. Indeed, typical simulation box snapshots as those shown in Fig. 2(b) clearly show that BS-Li-8, BS-Li-12, and BS-Li-16 form channel like structures by coordinating three anions with one Li ion. This can be precisely demonstrated by deriving the coordination numbers with the usual procedure of integrating the partial Li–S RDF over distances ranging from zero up to the (cutoff) position of the minimum following the distribution's main peak.

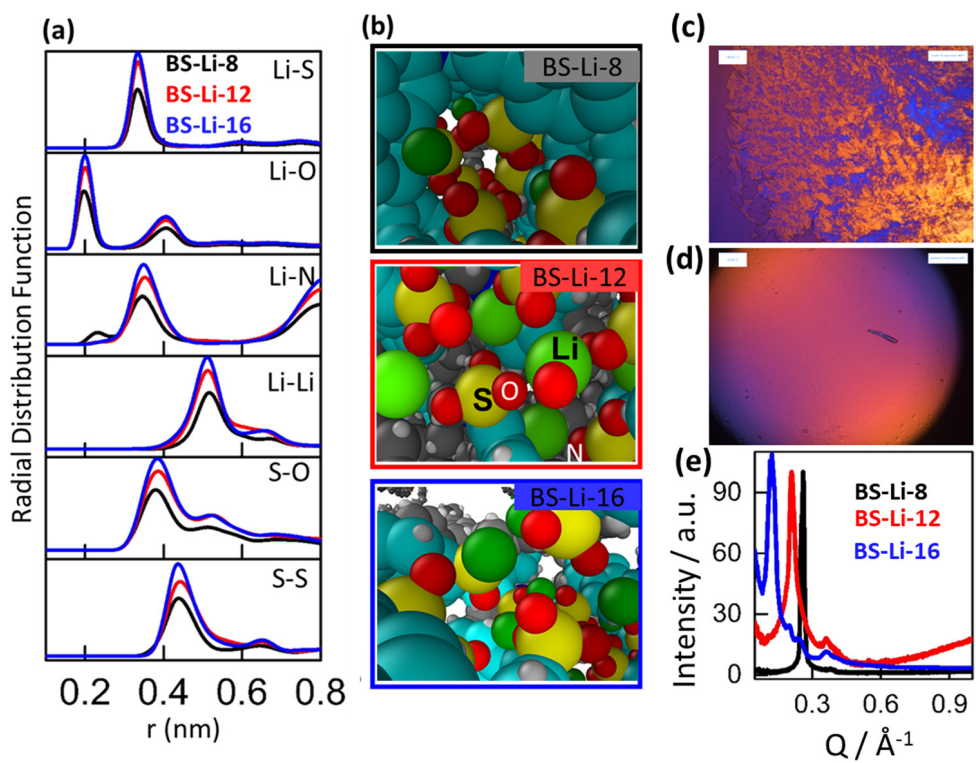

Fig. 2 Simulated radial distribution functions between Li–S, Li–O, Li–N, Li–Li, S–O, and S–S in (a) BS-Li-8 (black), BS-Li-12 (red), and BS-Li-16 (blue) at a temperature of 413 K, and simulated snapshots of the distribution of the LC molecules (b). Here nitrogen, benzene rings, carbon, hydrogen, sulphur, oxygen and Li ions are blue, cyan, gray, white, yellow, red and green, respectively. Polarizing optical microscopy images of LC molecules before (c) and after (d) rising temperature. (e) Small angle X-ray scattering of BS-Li-8 (black), BS-Li-12 (red), and BS-Li-16 (blue) under ambient conditions.

Additional structural information can be extracted from the simulated static structure factor, $S(Q)$. In Fig. S10 (ESI†), we show the $S(Q)$ of BS-Li-8, BS-Li-12, and BS-Li-16, at temperatures $T$ = 353, 373, 393 and 413 K. Although the main diffraction peaks are quite broad, we can identify the highest $S(Q)$ intensities for BS-Li-8, BS-Li-12, and BS-Li-16 at the non-monotonously varying positions 0.30, 0.32 and 0.26 $\text{Å}^{-1}$, respectively. With the exception of a relatively small decrease in intensity, we also found that modifying the temperature has little effect on the peak position and width in all cases. Also, from the presence of a sharp peak at large values of $Q$ in BS-Li-12, we can assume that these molecules are characterized by a higher degree of molecular aggregation than in the other two cases.

The above structural information extracted from MD simulation has an obvious drawback. Indeed, due to the finite size of our simulation boxes, limited to a few nanometers, we cannot extract sensible information on the large-scale (small-$Q$) assembly of our phases to complete the above nano/meso morphological picture. We have mitigated this limitation by probing the conformation of all materials in experiments.

We examined the temperature dependence of the structural behaviour of the ionic compounds by polarized light optical microscopy (POM, see above). We prepared our samples by shearing the materials between two glass plates throughout a temperature increase from ambient temperature to above 210 °C. We have found that the compounds organize into a macroscopic, rotationally homogeneous distribution of LC domains, as demonstrated by the presence of birefringence in the micrograph of Fig. 2(c). In contrast, POM micrographs, as shown in Fig. 2(d), allowed us to determine an isotropic state of the LC above 210 °C without any molecular degradation, as confirmed by TGA (Fig S6, ESI†). Interestingly, following a subsequent temperature decrease from the isotropic phase, we detected completely organized LC structures, without any defect textures, as confirmed by the absence of any sign of birefringence. The POM data overall confirm the ionic-surfactant-like behaviour of our materials: these are formed by molecules comprising an ionic head, and a non-polar moiety (naphthalene and alkyl chains), which leads to the nano-segregation of the polar network from the hydrophobic moieties, with the formation of layered phases with intercalated hydrophobic and ionic layers.

More quantitative information can be extracted from the SAXS patterns in Fig. 2(e), where BS-8-Li, BS-12-Li, and BS-16-Li show first order diffraction peaks at $Q$ = 0.26, 0.21, and 0.13 $\text{Å}^{-1}$, respectively, indicating the long-range ordering of the crystallized ionic phase. The position of the second peak observed on the SAXS patterns is at roughly twice that of the first peak. This suggests that it is consistent with a lamellar phase of thickness $2\pi/Q$, in the tens of nm, which increases with the length of the alkyl chains. The size of this larger structural feature is beyond the reach of MD simulations.

### 4.2 Transport

The ionic conductivities of all electrolytes were calculated from the MD simulation runs in terms of the computed Onsager transport coefficients,[46] $L^{ij}$, which fully describe the effect of correlations on the motion of particle types i and j. We employed the OCTP plugin to calculate the Onsager coefficients.[47] In the case of our LC electrolytes, three correlations are significant, e.g., cation–cation ($L^{++}$), anion–anion ($L^{--}$), and cation–anion ($L^{+-}$), respectively. $L^{++}$, $L^{--}$, and $L^{+-}$ for BS-Li-8, BS-Li-12, and BS-Li-16 at temperatures of 353 K to 413 K are shown in Fig. S11(a)–(c) (ESI†).

Some observations are in order. We can expect that two different cations should move in an anti-correlated fashion due to the mutual repulsion, therefore providing a negative contribution to $L^{++}$. The self-correlated terms, however, or those where the considered cations are part of the same aggregate and move together, in fact provide positive contributions to $L^{++}$, whose total value, we recall, must always be positive. Note that the anion–anion transport coefficient $L^{--}$ follows the same trends as $L^{++}$, as negative charges are strongly aggregated with the cations, as described above. Interestingly, $L^{+-}$ shows positive values lower than the other two coefficients, although in a standard single ion conductive electrolyte with weak cation anion interactions $L^{+-}$ could be negative.

In a system like this one, where both the ion and the counterion macromolecule are mobile, the choice of the reference frame plays a role in the transport properties. Our calculation systematically removed any residual center of mass velocity in the course of the trajectory. In ionic liquids, momentum conservation in this case imposes some constraints between the different Onsager coefficients.[48] However, although the alkyl chains are attached to the anions, they are long and flexible so the correlation between the motion of the anion and the atoms in its alkyl tails can be inaccurate at the time scales afforded by the simulation. One would need to carry out simulations reaching anionic displacement distances of several unit cells, in order to accurately account for this correlation, which is beyond our current capabilities. Therefore, at not overly long time scales, the flexible alkyl chains are always able to compensate any change of momentum of cations and anions without breaking off the latter. The results are shown in the ESI.†

We now focus on the ionic conductivity, $\sigma$, as shown in Fig. 3(a) of all materials, as a function of temperature. $\sigma$ can be calculated from the computed Onsager transport coefficients as[49]

$$\sigma = F^2(z_+^2 L^{++} + z_-^2 L^{--} + 2 z_+ z_- L^{+-}) \qquad (1)$$

Here, $F$ is the Faraday constant, $z_+$ and $z_-$ are the charge valences of the cation and anion, respectively, $L^{++}$, $L^{--}$, and $L^{+-}$ are the Onsager coefficients obtained from the simulation,

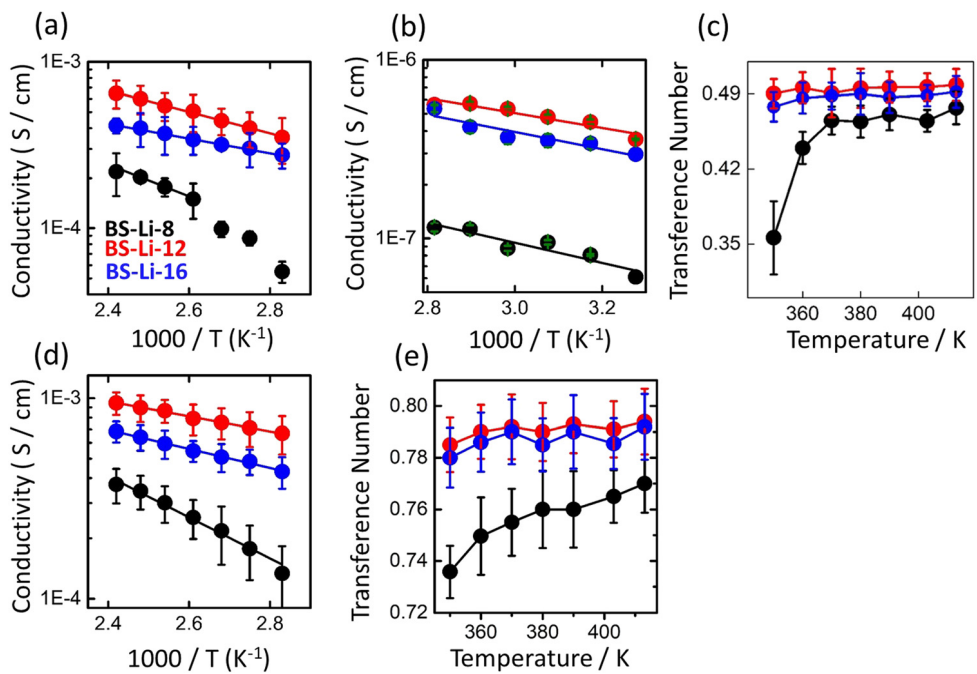

Fig. 3 (a) Simulated conductivity calculated using Onsager's coefficients, (b) experimental conductivity, (c) simulated Li$^+$ transference number calculated from Onsager's coefficients, (d) simulated conductivity and (e) transference number calculated using the Nernst–Einstein equation of BS-Li-8 (black), BS-Li-12 (red), and BS-Li-16 (blue) as a function of temperature.

as discussed above. These data interestingly show that, contrary to what one could naively expect, the ionic conductivity does not depend monotonously on the alkyl chain lengths, at any temperature. In particular, BS-Li-12 reaches the highest conductivity of about $6.4 \times 10^{-4}$ S cm$^{-1}$ at 413 K, a value which is approximately a factor of 3 and 3/2 higher than those pertaining to BS-Li-8 and BS-Li-16, respectively, at the same temperature.

Also, in BS-Li-8, the simulated conductivity decreases drastically below 373 K. As a consequence, while in the former cases the temperature behavior of $\sigma$ can be convincingly described by a single Arrhenius equation in the entire temperature range (solid lines in Fig. 3(a)), in the latter a cross-over between two Arrhenius regimes with different activation energies (higher at lower temperature) is visible. The details of the Li$^+$ ion conduction mechanism we can deduce from these data will be discussed below.

We now focus on the transference number, *i.e.*, the ratio of the current contribution due to the cation over the total ionic current. It is well known that it is preferable for an electrolyte to exhibit a high cation transference number (close to unity) to minimize anionic current contributions, together with the associated formation of detrimental concentration gradients.[49]

The Li$^+$ transference number, $t_+$, can, again, be expressed in terms of the computed Onsager coefficients as[49]

$$t_+ = \frac{z_+^2 L^{++} + z_+ z_- L^{+-}}{z_+^2 L^{++} + z_-^2 L^{--} + 2 z_+ z_- L^{+-}} \quad (2)$$

and the data are shown in Fig. 3(c). We first note that in the cases of BS-Li-12 and BS-Li-16, the simulated $t_+$ is almost temperature independent with $t_+$ of about 0.49, while for BS-Li-8 it follows a more complex behavior, with a constant value of $t_+ = 0.48$ at high temperatures crossing-over to a substantial decrease with temperature at low $T$, reaching values as low as $t_+ = 0.35$. Second, we observe that the transference numbers of our LC electrolyte simulations are overall substantially smaller than the nearly ideal value (about 0.9) reported in experiments for single ion conducting electrolytes.[50] We can rationalize this discrepancy by noticing that, due to the challenge posed by a direct measurement, many experiments determine the transference number approximately *via* the Nernst–Einstein expression, which considers an ideal uncorrelated transport of ions, at variance with our calculation based on eqn (4). This hypothesis is obviously hardly fulfilled by our highly charged materials, and is alone sufficient to justify the inconsistency.[51]

In order to make the above point clearer, we have calculated $\sigma$ and $t_+$ again, from the MD simulation, but now using the Nernst–Einstein approximation prescriptions,[49]

$$\sigma^{NE} = \frac{F^2}{RT}(z_+^2 D_+ + z_-^2 D_-) \quad (3)$$

$$t_+^{NE} = \frac{z_+ D_+}{z_+ D_+ - z_- D_-} \quad (4)$$

Here, $R$ is the gas constant, $T$ is the temperature, and $D_+$ and $D_-$ are the diffusion coefficients of the cation and anion, respectively. We show our data in Fig. 3(d) and (e). As expected, the measured conductivity $\sigma^{NE}$ is now consistently higher than that previously described, in all cases, while the general picture is unchanged. The $t_+^{NE}$ values, in addition, are substantially higher than those determined by correctly taking into account ionic correlations, assuming values in the range 0.74 to 0.8, which are now closer to the single-ion conducting ones.

We have complemented the numerical data with the experimental conductivity results, as shown in Fig. 3(b). These data point to an overall picture similar to that established by simulations. The two sets of data do not cover the same temperature range, since at lower temperatures the ionic hopping events in the MD simulation become too rare to reliably determine diffusion coefficients. The MD results are significantly higher than those found in the experiments. This is an expected limitation of the MD simulations that we discuss in the next section below. We found that BS-Li-12 again shows the potentially highest conductivity, followed by 16 and 8.

## 5. Discussion

We now provide our overall understanding of the data described above, starting with an observation. A fair hypothesis to explain the difference in the numerical values of the simulated and the experimental conductivity, beyond the obvious limitations of our classical, non-polarizable description of the LC, is the presence of domain boundaries in the bulk of the real materials, absent in the MD samples. As discussed earlier, POM suggests the presence of a distribution of finite size lamellar domains randomly oriented in space, implying an isotropic orientation of the ionic nanochannels. As a consequence, domain boundaries are present, which limits the diffusion of the Li$^+$ cation at the interfaces, thus substantially degrading the overall bulk ionic conductivity. It is well known that grain boundaries in other materials can lower the intrinsic ionic conductivity of solid electrolyte materials by orders of magnitude.[52–54] The immediate implication of the last observation is that in order to optimize the conductivity performance one should (a) minimize the presence of domain boundaries, controlling the orientational distribution of the ionic domains and maximizing the inter-channel ionic exchanges; and/or (b), most important here, clarify the intra-channel ionic transport mechanism. For the latter, we can reasonably rely on our simulation data.

We start by quantifying the typical size of the formed ionic domains at all investigated temperatures. The average size of the channels can be estimated by partitioning the S-atoms in disconnected clusters, defined as the subsets of atoms each of which is within the cutoff distance from one or more other atoms in the cluster.[55] The algorithm implies the use of a cutoff distance that we choose as the position of the first minimum following the main peak of the partial S–S RDF. By averaging over the entire MD trajectory, we can therefore extract the distribution of the cluster sizes, determined simply by counting the number of S-atoms in each cluster. (Note that this is not the spatial size.) In Fig. 4(a)–(c), we show typical clusters present in our snapshots for all materials, at 413 K. We have been able to identify 7, 4, and 5 clusters in BS-Li-8, BS-Li-12, and BS-Li-16, respectively, with typical life-times of the order of the overall simulation time. The temperature variation of both the average cluster sizes and the corresponding standard deviations (normalized to the average) are shown in Fig. 4(d). We immediately see that the largest clusters are found in BS-Li-12, followed by BS-Li-16, and finally BS-Li-8, therefore mirroring the trend of the ionic conductivity. In contrast, temperature does not significantly impact the typical average size of the ionic domains, which varies very mildly in the investigated temperature range. Interestingly, we found a substantial increase of size fluctuations by decreasing the temperature for BS-Li-12, while in the other two cases the standard deviations remain constant. Overall, the mere extension of the ionic domains seems to correlate with the relative transport efficiency of the materials, but not to account for the observed thermal behavior of the conductivity. The latter must be therefore related to the features of Li$^+$ transport (transfer) at a single-channel level.

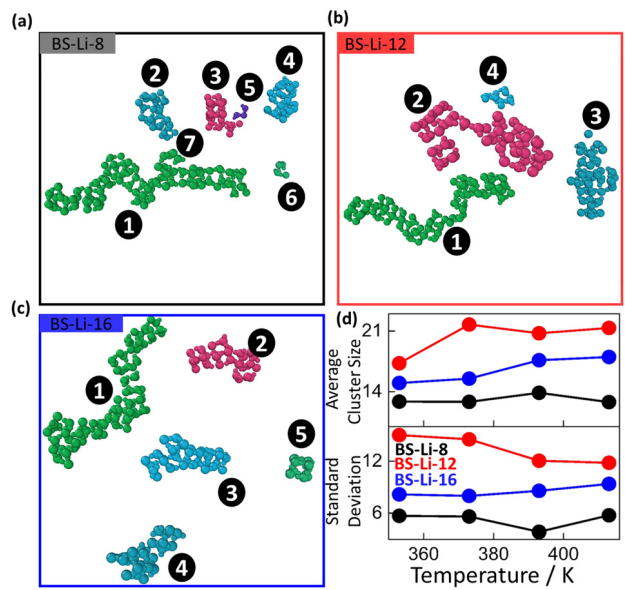

Fig. 4 Snapshots of the clusters formed by BS-Li-8 (a), BS-Li-12 (b), and BS-Li-16 (c) extracted from the 10 ns MD trajectories at 413 K, after 30 ns equilibration. Alkyl chains and benzene rings were omitted for clear visibility. (d) Average cluster number and corresponding standard deviation over the 10 ns trajectories, as a function of temperature, for BS-Li-8 (black), BS-Li-12 (red), and BS-Li-16 (blue).

We address this point by calculating the average residence time of the Li$^+$ ion at an ionic center as a function of temperature. This can be estimated by calculating the average time travelled together by a pair of neighboring species before separating, in our case Li$^+$ and S of the anion.[56] Our results are shown in Fig. 5(a). We obtain a picture perfectly consistent with our conductivity data, with longer residence times corresponding to lower values of $\sigma$, the same non-monotonic behavior with the alkyl chain length, and an increase with the decreasing temperature which correlates with the opposite behavior shown in Fig. 3. From these data, we conclude that both the multiscale organization of the ionic channels, controlled in an unexpected non-monotonic fashion by the length of the alkyl chains, and the strength of the interaction of the adsorbed Li$^+$

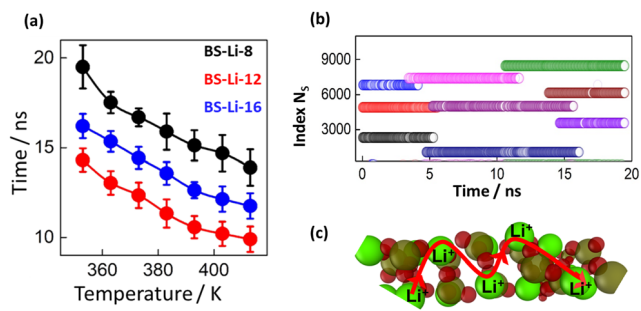

**Fig. 5** (a) Average residence time of Li$^+$ in the anionic center of BS-Li-8 (black), BS-Li-12 (red), and BS-Li-16 (blue). (b) Change of the S index of one Li$^+$ as a function of time in BS-Li-12. Index number of S (circle) and (c) simulated illustration of the possible Li$^+$ conduction route. Alkyl chains of LC molecules were omitted for better clarity.

ions at the anionic functional groups, concur to the ionic conductivity. Wisely modifying the chemical nature of the anionic center therefore turns out to be an additional viable option to optimize the conductivity performance of these materials.

We conclude by providing a few hints about the nature of the (completely classical) nano-scale Li$^+$ transfer in an ionic channel. In our simulations, we have followed the single Li$^+$ ions, monitoring at high frequency the interacting sulfonic groups, on very long time scales. This amounts not only to determine the fluctuations in time of the total cations–anion coordination number, but also to keep track of the identity of the interacting ionic centers. In Fig. 5(b), we show a typical data set extracted from this procedure, where at each time step the indexes of the instantaneously interacting S-atoms are indicated by circles of different colors. We recall that, on average, in all materials and at all temperatures, each Li$^+$ ion coordinates with 3 sulfonic groups. By visually inspecting these data, we realize that the total coordination number varies intermittently, jumping back and forth from the state with coordination 3, occupied for very long times, through short-lived transitions to higher values (4 or even 5). Interestingly, at the end of each of these cycles, the identity of all three interacting anions has systematically changed, signaling the actual transfer of the ion on length scales of the order of a few anions spacing, as depicted in the cartoon of Fig. 5(c). The main transport mechanism therefore turns out to be a quite effective sliding of the cation along the ionic channels, mediated by fast transitions to overcrowded highly-negatively charged local environments. Similar to BS-Li-12, the other LC samples also follow this transport mechanism (Fig. S12 and S13, ESI†).

## 6. Conclusions

We have synthesized and analyzed a novel liquid crystal electrolyte composed of rigid naphthalene-based mesogenic units, attached to flexible alkyl chains of different lengths. By using molecular dynamics simulations and advanced experimental techniques, we have fully characterized both phase properties and transport behaviour, together with their temperature modification and alkyl chain length variations. Our main findings include the following points. First, SAXS and optical microscopy have allowed us to demonstrate the formation of ordered ionic domains with lamellar symmetry, characterized by correlation distances increasing with the length of the aliphatic chains. The MD simulation, in addition, has provided specific information about the local nanostructure of the compounds, clarifying both the details of the cation/anion interaction and the coordination properties. Next, we have aimed at a complete characterization of the ionic transport features. By using both numerical data and electrochemical impedance spectroscopy, we have demonstrated an intriguing non-monotonous dependence of the ionic conductivity with the alkyl chain length, with lithium 4 – (didodecylamino) naphthalene – 1 sulfonate showing the best performance over the other materials, in the entire considered temperature range. We have tentatively associated this behaviour with a more efficient formation of the ionic domain network in this case whose origin, however, is still not completely clarified. Finally, based on our simulations, we have rationalized our findings in terms of a clearly identified lithium hopping process at the nanoscale, already described in previous work on similar materials and alternative to other efficient conduction mechanisms, like those assisted by different forms of segmental motion. Overall, on the one hand, we have demonstrated that materials based on the molecular concept we have considered here show transport features that can make them promising electrolyte candidates for all-solid-state Li-ion battery applications. On the other hand, we have provided new insight for a more efficient navigation of the extremely vast design space of this class of materials.

## Conflicts of interest

There are no conflicts of interest to declare.

## Acknowledgements

This work is supported by the Institute Carnot, under project PREDICT. S. M. acknowledges support by the project Move-YourIon (ANR-19-CE06/0025) funded by the French ''Agence Nationale de la Recherche''.